\documentclass[%
 aip,
  pop, % Physics of Plasmas substyle of AIP, no titles in citations
 amsmath,amssymb,
 preprint,%
]{revtex4-1}
\usepackage{CJK}

\usepackage{ulem}
\usepackage[T1]{fontenc}
\usepackage{calligra}

\usepackage{graphicx}% Include figure files
\usepackage{dcolumn}% Align table columns on decimal point
\usepackage{bm}% bold math

\usepackage{times}
\usepackage{verbatim}
\usepackage{graphicx}
\usepackage{graphics,epsfig}

\usepackage{theorem}
\usepackage{makeidx}
\usepackage{amsmath}
\usepackage{epic}
\usepackage{amscd}

\usepackage{bbm}
\usepackage{xy}
\usepackage{amssymb}
\usepackage{epsfig}
\def \phi {\mbox{$\varphi$}}

\begin{document}

\begin{CJK*}{GB}{}
\begin{CJK*}{KS}{}

\title{Continuum limit of electrostatic gyrokinetic absolute equilibrium}
\author{Jian-Zhou Zhu}
%\author{Jian-Zhou Jason Zhu (ñ¹Ëïñ¶,ÁÖ°ÅÁÖ)}
\affiliation{Department of Modern Physics, University of Science and Technology of China, Hefei, Anhui 230026 China}

\begin{abstract}
Electrostatic gyrokinetic absolute equilibria with continuum velocity field are obtained through the partition function and through the Green function of the functional integral. The new results justify and explain the prescription for quantization/discretization or taking the continuum limit of velocity. The mistakes in the Appendix D of our earlier work [J.-Z. Zhu and G. W. Hammett, Phys. Plasmas {\bf 17}, 122307 (2010)] are explained and corrected. If the lattice spacing for discretizing velocity is big enough, all the invariants could concentrate at the lowest Fourier modes in a negative-temperature state, which might indicate a possible variation of the dual cascade picture in 2D plasma turbulence.

\end{abstract}

%\pacs{05.40.-a, 47.52.+j, 51.10.+y}

\maketitle

\section{Introduction}
The statistical equilibrium states of a Galerkin truncated system are called the absolute equilibria. Such equilibria have been used to study the statistical dynamics of various systems since Lee,\cite{Lee1952} including models for the Bose-Einstein condensates and superflows (see Ref. \onlinecite{KrstulovicBrachet2011} and references therein,) besides other systems such as classical (magneto)hydrodynamics, drift waves and (gyro)kinetic fluctuations in plasmas (see Ref. \onlinecite{ZhuHammettPoP2010} and references therein), among others. The novel feature in (gyro)kinetics is the appearance of the extra velocity variable in distribution functions. Different treatments of the extra dimension of the velocity variable would lead to different absolute equilibria. This paper studies gyrokinetics statistical equilibria in the case without additional treatment of the velocity variable. This system is thus closer to the original one. Our focus is on the general feature of the problem and the specific solving techniques, with discussions of the results in the common context of turbulence physics. As in the article by Krommes,\cite{Krommes12} the results presented here are of potential interest also to non-plasma physicists.
%\cite{FTref}
%\footnote{
(A minimization of references is thus managed to keep the general readers in an easy environment.)
%}
 This work is self-contained and, in particular, knowledge of the results in the paper by Zhu and Hammett \cite{ZhuHammettPoP2010} (henceforth referred to as I) is not required. As supplementary material to further highlight discretization schemes, we will actually provide an alternative approach, the partition function calculation, whose intermediate steps reproduce the results in I.

Zhu and Hammett \cite{ZhuHammettPoP2010} obtained gyrokinetic absolute equilibria which can be considered as an approximation of the ``exact'' continuum result. It is fundamental to ask what the continuum limits or the exact results are and how the continuum limits should be taken (this latter issue should have been clear from the discretization scheme, but there are subtleties). A deeper understanding of gyrokinetic absolute equilibrium is promising for plasma turbulence, as indicated by theoretical work in hydrodynamics. For example, Ref. \onlinecite{LPPprl02} found that in some cases two dimensional (2D) turbulence is actually not far from absolute equilibrium, and dissipative three dimensional (3D) turbulence is also partially thermalized (see, {\it e.g.}, Refs. \onlinecite{Lee1952} and \onlinecite{FrischPRL08}.)

As we will see, the problem itself is formulated in terms of integrals over the velocity field. Therefore, alternatively, we can perform the calculations without starting with the discretization of velocity used in I, and the results should be the continuum limits I. This is the main technical contribution of this paper, as detailed in Sec. \ref{calculation}. In other words, instead of first applying discretization and then taking the continuum limit, we will take the strategy of starting from the calculation with continuum velocity and then discretize the results. Such an approach provides an alternative view of the problem and may clarify the issue better: Suppose we have somehow arrived at the correct continuum limit, then figuring out the continuum limit directly will of course justify and explain the discretization/quantization scheme and the prescription of taking the continuum limit from the discretized results. Also, calculations with arbitrary velocity field offer direct comparisons and allow us to elucidate other issues such as the physical effects of numerical errors and of finite truncation of the velocity scale (typically from the collision operator). As explicitly stated in the Footnotes 46 and 65 of I, the coauthors disagree on what the continuum limit should be and on how to obtain the limit in the 2D case. To avoid confusion and extra burden of references to the general audience, we make this paper sufficiently self-contained. We clearly, though briefly, point out the mistakes in Appendix D of I. Together with the discretized-velocity results, the continuum limits may also promote more fundamental considerations on the statistical dynamics of magnetized plasmas.

\section{The Model}
\label{model}
Gyrokinetics is a powerful model for modern research on microturbulence in magnetized plasmas with many applications to both astrophysics and fusion (see, e.g., Krommes \cite{Krommes12} - which is intended also for non-plasma physicists.) Here, we limit ourselves to the nonlinear electrostatic gyrokinetic Vlasov-Poisson equations as in I, where more literature can be found.

The gyrokinetic equation for electrostatic fluctuations in slab geometry with uniform background magnetic field $\textbf{B}_0=B_0\textbf{z}$ reads
\begin{eqnarray}\label{eq:gyronorm}
\frac{\partial g}{\partial t}
+v_\parallel \frac{\partial g}{\partial z}
+ \left( \hat{\textbf{z}} \times \frac{\partial \langle \varphi
  \rangle_{\textbf{R}} }{\partial \textbf{R}}
\right) \cdot \frac{\partial g}{\partial \textbf{R}}
= - v_\parallel
\frac{\partial \langle \varphi \rangle_{\textbf{R}}}{\partial z} F_{0},
\end{eqnarray}
where we follow the notations and normalization conventions used earlier in I: $g(\textbf{R},v_{\parallel},v_{\perp})$ defined at the gyrocenter is a component of the fluctuating distribution $f$, where $f=\exp\{-q\varphi\}F_0+g+\langle \varphi \rangle_{\textbf{R}}+h.o.t.$, which gives the deviation from the maxwellian $F_0$. Here $h.o.t.$ denotes the higher-order terms in the gyrokinetic ordering and $\langle \cdot \rangle_{\textbf{R}}$ the gyroaverage around $\textbf{R}$, i.e., the average of any quantity $\Psi$ along a ring of gyroradius $\rho$ surrounding the guiding center $\textbf{R}$
and perpendicular ($\perp$) to the magnetic field direction ($\parallel$): $$\langle \Psi \rangle_{\textbf{R}}=\frac{ \int \Psi(\textbf{r}) \delta( \textbf{r}_{\parallel} - \textbf{R}_{\parallel} ) \delta[ |\textbf{r}_{\perp}-\textbf{R}_{\perp}|-\rho(\textbf{R}) ] d^{3}\textbf{r} } {\int \delta( \textbf{r}_{\parallel} - \textbf{R}_{\parallel} ) \delta[ |\textbf{r}_{\perp}-\textbf{R}_{\perp}|-\rho(\textbf{R}) ] d^{3}\textbf{r}}$$
with
%$\boldsymbol{\rho}=\textbf{x}-\textbf{R}$ and
%$q\boldsymbol{\rho}=m\textbf{v}_{\perp}\times \textbf{B}/B^2$
$q\boldsymbol{\rho}\times \textbf{B}=m\textbf{v}_{\perp}$, which illustrates how $\boldsymbol{\rho}$ and $\textbf{v}_{\perp}$ experience the gyroangle together. [Throughout this article, unless explicitly specified, $\int$ is used to denote the definite integral over the entire domain. Using a Fourier representation $\Psi(\textbf{r}) = \sum_{\textbf{k}}
\exp(-i\textbf{k} \cdot \textbf{r}) \hat{\Psi}_\textbf{k}$, and considering a
straight magnetic field for the sake of simplicity, we have $\langle \Psi
\rangle_{\textbf{R}}=\sum_\textbf{k} \exp(-i \textbf{k} \cdot
\textbf{R}) J_0(k_\perp \rho) \hat{\Psi}_{\textbf{k}}$, where $J_0$ is a Bessel
function.\cite{Smagorinsky63}
%\footnote{The explicit spacial integration form however reminds us the large eddy simulation proposed by Smagorinsky \cite{Smagorinsky63}.}
] The gyrokinetic equation expresses how guiding centers evolve in
time due to parallel motion along the magnetic field, the gyro-averaged
$\textbf{E} \times \textbf{B}$ drift across the magnetic field (this is
the nonlinear term), and the parallel electric field acceleration.
This equation is closed by using the gyrokinetic quasi-neutrality
equation to determine the electrostatic potential, which in Fourier
space in normalized units is given by
\begin{eqnarray*}\label{eqQN3D}
\hat{\varphi}(\textbf{k},t) = \frac{\beta(\textbf{k})}{2 \pi}
   \int d^3v J_0(k_{\perp}  v_{\perp})
   \hat{g}%\nonumber \\%(\textbf{k},v_{\parallel},v_{\perp},t) \nonumber \\
  = \beta(\textbf{k}) \! \int_{-\infty}^\infty \! \! dv_{\parallel}
   \int_0^\infty \! \! dv_{\perp} v_{\perp}
   J_0(k_{\perp}  v_{\perp})
   \hat{g}. %(\textbf{k},v_{\parallel},v_{\perp},t), \nonumber \\
\end{eqnarray*}
The physical (dimensional) variables having subscript `p' are normalized as follows:
$t = t_{\mbox{\scriptsize{p}}}v_{\mbox{\scriptsize{th}}}/L$, $x = x_{\mbox{\scriptsize{p}}}/\rho_{th}$, $y = y_{\mbox{\scriptsize{p}}}/\rho_{th}$, $z = z_{\mbox{\scriptsize{p}}}/L$, $v = v_{\mbox{\scriptsize{p}}}/v_{\mbox{\scriptsize{th}}}$, $\varphi = \varphi_{\mbox{\scriptsize{p}}}\frac{q L}{T_0 \rho}$, $h = h_{\mbox{\scriptsize{p}}}\frac{v_{\mbox{\scriptsize{th}}}^3 L}{n_0 \rho}$ and $F_0 = F_{0\mbox{\scriptsize{p}}}v_{\mbox{\scriptsize{th}}}^3/n_0$.
The equilibrium density and temperature of the species of interest are $n_0$ and $T_0$; the thermal velocity is $v_{\mbox{\scriptsize{th}}} = \sqrt{T_0/m}$; the Larmor radius is $\rho_{th} = v_{\mbox{\scriptsize{th}}}/\Omega_{c}$ where the Larmor (cyclotron) frequency is $\Omega_{c} = qB/m$.  $L$ is the reference scale length (\textit{i.e.}, the system size), satisfying $\rho/L \ll 1$ for consistency with the gyrokinetic ordering.

For plasmas in a two dimensional (2D) cyclic box, with $v_{\parallel}$ being integrated out and the subscript $_{\perp}$ in $v_{\perp}$ omitted, the whole system in wavenumber space is
\begin{eqnarray}\label{eq:2DgkK}
\partial_t \hat{g}(\textbf{k},v)-\textbf{z}\times \sum_{\textbf{p}+\textbf{q}=\textbf{k}} \textbf{p}J_0(pv)\hat{\varphi}(\textbf{p})\cdot \textbf{q} \hat{g}(\textbf{q},v)=0
\end{eqnarray} and
\begin{eqnarray}\label{eqQNK}
\hat{\varphi}(\textbf{k})=\beta(k) \int vdv J_0(kv) \hat{g}(\textbf{k},v).
\end{eqnarray}
Here $\beta(k)=\frac{2\pi}{\tau+1-\hat{\Gamma}(k)}$ and $\hat{\Gamma}(x)=I_0(x^2)e^{-x^2}$ is an exponentially-scaled
modified Bessel function, where $I_0(x) = J_0(i x)$
and $\tau$ represents the shielding by the other species which is treated as
having a Boltzmann response of some form (an anisotropic response model, such as those respecting the zonal modes, can also be considered as in I).

We should remark that the electrostatic gyrokinetic equations constitute an integro-differential system. When they are discretized, we denote the corresponding solution by $\tilde{g}$ to distinguish it from the exact solution $g$ of the original equations. In general, $\tilde{g}(v)\neq g(v)$. As our interests are in the statistical properties of the Fourier Galerkin truncated system for which $g$ would be reserved in the case with continuum velocity, $\tilde{g}$ is used for the solution to the system with further discretization of the velocity. To our best knowledge, there is no rigorous mathematical result on the singularities yet, not to mention the collisionless dissipation of the gyrokinetic equations. However, we believe the Fourier Galerkin truncation will also smooth the velocity space structure and $g$ here should not produce dissipation, unlike the shock solutions in Burgers equation, so that it should make sense to talk about a statistical equilibrium state.

\section{Calculations}
\label{calculation}

The Fourier Galerkin truncation here is defined by setting all Fourier modes beyond the wave number set $\mathbb{K}=\{\textbf{k}:k_{min}<|k|<k_{max}\}$ (the summation over which will be denoted by $\tilde{\sum}$) to zero. As Lee \cite{Lee1952} did for Euler equation for ideal fluid flow, we first observe that the dynamics of the ``gas'', composed of the real and imaginary parts of the Fourier modes (denoted by $\sigma$), satisfies Liouville theorem. Actually ${\delta\dot{\sigma}\over \delta\sigma}=0$, where the dot represents time derivative. Then, we proceed to find the canonical distribution. Let us start with the 2D case. The only known rugged (conserved after Fourier Galerkin truncation) invariants for the 2D gyrokinetic system are
$$G(v)=\int \frac{d^2\textbf{R}}{2\mathcal{V}} g^2=\tilde{\sum}\frac{1}{2}|\hat{g}(\textbf{k},v)|^2$$ and the mean effective electrostatic potential ``energy'' $$E=\int \frac{d^2\textbf{r}}{2\mathcal{V}} [(1+\tau) \varphi^2-\varphi \Gamma \varphi]=\tilde{\sum}\frac{\pi}{2\beta(\textbf{k})} \left| \hat{\varphi}(\textbf{k}) \right|^2,$$ (c.f., e.g., I and references therein) where $\mathcal{V}$ is the volume (area) of the integration domain. Notice that $G(v)$ is a function of $v$, so we have now ``one plus a continuum'' of conserved quantities.

\subsection{Green function}
The Gibbs distribution\cite{EscandeJSP94} reads
%\footnote{There does not seem to be a complete theory for applying such a distribution (but see, for example, Escande \cite{EscandeJSP94} for a validity check of the Gibbs technique).}
$$Z^{-1} \exp\{-\mathcal{S}\},~Z=\int \!\! D\sigma  \exp\{-\mathcal{S}\},$$ with $$\mathcal{S}=\int \alpha(v)\mathrm{G}(v)dv+\alpha_0\mathrm{E}$$ and with $\sigma$ running over the configuration space spanned by the real and imaginary parts of $\hat{g}$,\cite{Lee1952} which we don't expose explicitly.
Here the constant of motion $\mathcal{S}$ is formed by introducing
$\alpha_0$ and $\alpha(v)$, the ``(inverse) temperature parameters'' as the Lagrange multipliers when  maximizing the Gibbs ensemble entropy:
\begin{eqnarray}\label{eq:H2Dc}
\!\!\!\!\!\!\!\mathcal{S} = \frac{1}{2} \tilde{\sum}_{\textbf{k}} \Bigg\{ \int \alpha(v) |\hat{g}(\textbf{k},v)|^2 dv+\alpha_0 2\pi  \beta(k) %\nonumber\\
\int vdv J_0( kv) \hat{g}(\textbf{k},v)\int vdv J_0( kv) \hat{g}^{*}(\textbf{k},v) \Bigg\}\nonumber\\
\!\!\!\!\!\!\!= \tilde{\sum}_{\textbf{p}}\tilde{\sum}_{\textbf{k}} \frac{1}{2} \int\int \hat{g}(\textbf{p},u) \mathbf{M}(\textbf{p},\textbf{k},u,v)\hat{g}^{*}(\textbf{k},v)du dv,
\end{eqnarray}
where ``$^{\ast}$'' denotes complex conjugate. Clearly, $\mathbf{M}(\textbf{p},\textbf{k},u,v)=\delta_{\textbf{p},\textbf{k}}C^{\mathcal{I}}(\textbf{k},u,v)$,
%\cite{FieldIntegral}
%\footnote{In quantum and statistical field theory \cite{FieldIntegral}, $C^{\mathcal{I}}$ here is called the {\it operator} kernel or {\it propagator} and $C$ below can be interpreted as the {\it Green function} of $C^{\mathcal{I}}$. [The measure $D\sigma$ is usually written as $D(\hat{g}^{\ast},\hat{g})$ with $\hat{g}$ being the function of $\textbf{k}$ and $x$. But, note that our stationary integration measure does not depend on time $t$.] Our case is just the functional Gaussian (but not Grassmann) integral, but the propagator presents new features which require novel techniques.}
with
\begin{eqnarray}\label{eq:CIuv}
C^{\mathcal{I}}(\textbf{k},u,v) = \alpha(v)\delta(u-v)+ \alpha_0 2\pi \beta(k) vuJ_0( kv)J_0( ku).
\end{eqnarray}
[We should clarify that $G(v)$ is introduced as a density over $v$ in the first line of Eq. (\ref{eq:H2Dc}) and is then changed to a density in the $u-v$ plane in the second line with the help of the Dirac delta function: Since $\langle |\hat{g}(\textbf{k},v)|^2 \rangle=\langle \hat{g}(\textbf{k},v)\hat{g}^*(\textbf{k},v) \rangle$, the variable $v$ can be regarded as appearing once or twice; so, a density of $v$, or, of $u$ and $v$ with $u=v$ (denoted as ``1V'' and ``2V"
respectively) could be represented depending on the context, which should be kept in mind in case of misconceptions; similarly is the discretized case.
%\cite{FTdv}
%\footnote{
(``D'' is already used to denote the dimension in configuration space, so we adopt ``V'' here for the dimension in velocity space.)
%}
The 1V and 2V densities, though written symbolically the same, acquire different values.]

With the definition of functional inverse,
\begin{eqnarray}\label{eq:CIC}
  \!\!\!\!\!\! \int \!\!C^{\mathcal{I}}(u,x)C(x,v)dx \!=\! \delta(u\!-\!v) \!=\!\! \int \! C(u,x)C^{\mathcal{I}}(x,v)dx,
\end{eqnarray}
we find (for details, see a simple approach with careful observations in Appendix \ref{APP:FunctionalInverseI} and a general functional inverse formula which leads to the same result as established in Appendix \ref{APP:FunctionalInverseII})
\begin{eqnarray}\label{eq:Cuv}
\!\!\!\!\!\! C(\!\textbf{k},\!u,\!v\!) \!\!=\!\! \frac{ \delta(\!u\!-\!v\!)}{\alpha(v)} \!\!-\!\! \frac{2\pi\alpha_0 \beta(k) u J_0( uk)v J_0( vk)} {\alpha(u)\alpha(v)\!\! \left[ \! 1\!\!+\!\!2\pi\alpha_0 \beta(k)\!\! \int \!\! {x^2 \! J_0^2( kx)\over \alpha(x)}dx \! \right]}.
\end{eqnarray}
Here
\begin{eqnarray}\label{eq:CorrelationDensity}
C(\textbf{k},u,v)=\langle \hat{g}(\textbf{k},u) \hat{g}^{\ast}(\textbf{k},v)\rangle,
\end{eqnarray}
where $\langle \cdot \rangle$ denotes the statistical average over the Gibbs distribution $\int D\sigma (\cdot)\exp\{-S\}/Z$
%\footnote{\label{realizability}
[A factor 2 has been eliminated to respect the realizability condition $g^*(\textbf{k},u)=g(-\textbf{k},u)$]
%}
%\cite{FTrealizability}
, is a 2V density in the $u$-$v$ plane. Therefore, given the Dirac delta function in the right hand side of Eq. (\ref{eq:H2Dc}),
the 1V distribution over $v$ of the spectra density (over $\textbf{k}$) of $g^2(\textbf{R},v)$ can be obtained as detailed in Appendix \ref{APP:2to1}:
\begin{eqnarray}\label{eq:Gvspectrum}
\textmd{G}(\textbf{k},v)=1/\alpha(v).
\end{eqnarray}
This spectrum is independent of $\textbf{k}$ (``equipartition'') since the second term of Eq. (\ref{eq:Cuv}) does not contribute to the line $u=v$: see Eqs. (\ref{eq:infinitesimal1D2D}) and (\ref{eq:int1D2D}) --- the global picture is that $\mathrm{G}(v)$, extended as 2V later to facilitate the calculation, was introduced in Eq. (\ref{eq:H2Dc}) as a 1V density to which Eq. (\ref{eq:Gvspectrum}) is reverted.
When the 1V density is extended to the $u$-$v$ plane, a Dirac delta function must be introduced as in Eqs. (\ref{eq:Cuv}) and (\ref{eq:CIuv}), and, of course the 2V density is then singular for $u=v$, unlike the 1V density. The 1V result in Eq. (\ref{eq:Gvspectrum}) is the same as the case $\alpha_0=0$, because the diagonal ($u=v$) contribution of the $\alpha_0$ term is zero (as the 2D Lebesgue measure of a line is zero) and the nondiagonal ($u\neq v$) contributions cancel among themselves. We then calculate the spectral density $D(\textbf{k})= \frac{\pi}{\beta(\textbf{k})} \langle \left| \hat{\varphi}(\textbf{k}) \right|^2 \rangle$ of the ``energy'' $E$ by
\begin{eqnarray}\label{eq:elecspecC}
D(\textbf{k}) = \pi \beta(k) \int \!\!\! \int C(\textbf{k},u,v)vJ_0( kv) uJ_0( ku) du dv%\nonumber\\
= \frac{\pi \beta(k) \int {v^2 J_0^2( kv)\over \alpha(v)}dv}{1+\alpha_0 2\pi \beta(k) \int {v^2 J_0^2( kv)\over \alpha(v)}dv}.
\end{eqnarray}

%If velocity is quantized and things like the temperatures are locally (in velocity coordinate) uniform so that a natural lattice appears, or if one wants to do simulation by computers the results should be discretized. The 2V convariance density $C$ in Eq. (\ref{eq:Cuv}) contains a
%Dirac delta distribution, which is defined through its integral. Thus, the proper way to discretize
%$C$ should be to match the behavior of the integral as our problem is formulated.
When discretizing, assuming rectangular lattice for simplicity, with particularly
\begin{eqnarray}\label{eq:delta}
\delta_{i,j}/m_i \leftrightarrows \delta(u-v)
\end{eqnarray}
we have the covariance density matrix elements
\begin{eqnarray}\label{eq:CuvD}
c_{i,j}(\textbf{k}) = \langle \tilde{\hat{g}}(\textbf{k},v_i)\tilde{\hat{g}}^{\ast}(\textbf{k},v_j)  \rangle %\nonumber\\
= \frac{\delta_{i,j}}{\alpha_i}-
  \frac{\tilde{\alpha}_0 2 \pi \beta(k) w_i
    \alpha_i^{-1}w_j\alpha_j^{-1}}{1+\tilde{\alpha}_02\pi \beta(k) \sum_l w_l^2
    \alpha_l^{-1}}
\end{eqnarray}
which is a 2V discrete density.
Here $w_i(k) = m_i v_i J_0( k v_i)$, and $m_i$ is the weight of
velocity grid point $v_i$. Actually, now $m_i
= \Delta v_i$; and,
\begin{eqnarray}\label{eq:alpha}
\alpha_i=\tilde{\alpha}(v_i)m_i, \ \text{for} \ i=1,...,N,
\end{eqnarray}
where $N$ is the total number of grid points.
The tilde above $\tilde{\alpha}$ is similar as for $\tilde{g}$. (Such discretization corresponds to schemes used by current Eulerian gyrokinetic codes and our thermodynamic limit is exactly taken with the lattice spacing going to zero.)
%\cite{FTdis}
%\footnote{
In I, the same covariance density matrix as in Eq. (\ref{eq:CuvD}) was obtained by using the discretization of Eqs. (\ref{eq:2DgkK}) and (\ref{eqQNK}) as the starting point. Especially, to facilitate the calculation, $\tilde{G}_i$ was multiplied by $\delta_{i,j}$, which, according to the current Eq. (\ref{eq:delta}), transformed the 1V density to 2V density: $\tilde{G}_i/m_i \Rightarrow G_i$, with the notations used there.
%}
Note that now $c_{i,j}$, as well as $\alpha_i$, depend on the size $\Delta v_i \times \Delta v_j$ of the 2V lattice into which the distribution curdling on the 1V line $u=v$ is dissolved. If a physical example is still required, we would mention that Mandelbrot \cite{MandelbrotJFM74} proposed that the turbulent dissipation of a fluid is curdling into fractals, which means a Dirac delta function supported by the fractal object (carrier) imbedded in the 3D space (or 4D space time.) The introduction of a lattice for discretization facilitates the computation of box counting dimension whose value is actually 1 for our case here (we have a full line, not really fractal.)
%\cite{FTfractal}
%\footnote{
%[Note however that we have not proposed here, in velocity space, the fractals which may be appropriate for some physical properties though. With the fractal as an example, it is indicated that the technique may be extended to cases with fractional dimensions.]
%}
Especially, we have the spectral density of the $(i,i)$th ``cell''
\begin{eqnarray}\label{eq:Gspec}
%\!\!\!\!\!\!\!\!\!\!\!\!\!\mathrm{G}_i(\textbf{k}) =
   c_{i,i}(\textbf{k}) = \langle \tilde{\hat{g}}(\textbf{k},v_i)\tilde{\hat{g}}^{\ast}(\textbf{k},v_i)  \rangle %\nonumber\\
   =
\frac{1}{ \alpha_i} \left[ 1 - \frac{\tilde{\alpha}_0 2\pi
  \beta(k) w_i^2 \alpha_i^{-1}}{1+\tilde{\alpha}_0 2\pi \beta(k)
  \sum_l w_l^2 \alpha_l^{-1}} \right]
\end{eqnarray}
which is a 2V density with the two velocity indexes coinciding. As mentioned, the continuum limit of $c_{i,j}$ contains a Dirac delta function which is singular for $u=v$, so this 2V density $c_{i,i}$ is singular in such a limit. The corresponding 1V density can be obtained by following the procedure given in Appendix \ref{APP:2to1}, only with the infinitesimals changed to be finite $\Delta u(v)$.
%\cite{FTgi}
%\footnote{
$\mathrm{G}_i$ in Eq. (11) of I is a 2V density so that, even though the factor of $1/2$ there is neglected, the summation over $\textbf{k}$ of it does not equal to $\tilde{G}_i$, which is 1V, in the left column on that same page.
%}
[$\mathrm{G}_i(\textbf{k})$ there should have been used to denote the 1V discrete density instead of the 2V density $c_{i,i}(\textbf{k})$ in Eq. (11) there. For this historical reason, we resist using $G_i$ in this paper.]

Note adding an arbitrary term, which vanishes as the lattice spacing goes to zero, will not change the convergence to the density [of Eq. (\ref{eq:CorrelationDensity}), or of Eq. (\ref{eq:Gvspectrum}) after transformed to the 1V density - see Appendix \ref{APP:2to1}] with continuum velocity. The additional term should be determined by other condition(s). In our case, the ``additional'' term comes from the two dimensional (covariant) density which is determined by the conserved quantities. For our 2V covariance function in continuum freedoms of $u$ and $v$, there is also a second term, the discretization of which exactly corresponds to the ``additional'' term for the discrete density, besides the one with the Dirac delta function. The effects of this ``additional" term as discussed in Sec. \ref{discussion} have been partly confirmed by ongoing numerical work.\cite{WatanabeZhu2011}
One should also be careful that $\langle \tilde{\hat{g}}(\textbf{k},v_i)\tilde{\hat{g}}^{\ast}(\textbf{k},v_i)\rangle = c_{i,i}(\textbf{k})$ here for the discretized version is not equivalent to Eq. (\ref{eq:CorrelationDensity}) evaluated at $u=v=v_i$, but to the coarse-graining of it [$\tilde{\hat{g}}(\textbf{k},v_i)$ is the solution to the discretized system but not $\hat{g}(\textbf{k},v)$ (the solution of the original system, evaluated at $v=v_i$)].
%%\cite{FTgki}
%%\footnote{\label{fn:gki}
%[So, to be more precise, for the discretized case earlier in I we should have replaced $\hat{g}(\textbf{k},v_i)$ with $\tilde{\hat{g}}(\textbf{k},v_i)$, denoted by $\hat{g}_{\textbf{k},i}$ for clarity, as is the case of distinguishing the solution of the discretized equation (say, the difference equation) of a differential equation and the solution of the latter.]
%%}

We thus have been able to go back and forth smoothly between the distribution over continuous $u,v$ (the zero lattice spacing limit) and the discretized version. Eqs. (\ref{eq:CuvD}) and (\ref{eq:Gspec}) are the same as those obtained earlier in I where the calculations were performed with the discretized system of Eqs. (\ref{eq:gyronorm}) and (\ref{eq:2DgkK}) as the starting point.
%Note again that $c_{i,j}$ is actually a 2V discrete distribution whose corresponding 1V discrete distribution can be easily obtained like that shown in \ref{APP:2to1} with only the infinitesimals $du(v)$ changed to finite $\Delta u(v)$.
%Without the present calculation with arbitrary (continuum) velocity field, it is however nontrivial to take the zero lattice size limit and misconceptions may arise: For example, without the understanding that, unlike Eq. (\ref{eq:Gvspectrum}), the left hand sides of Eqs. (\ref{eq:Gspec}), (\ref{eq:CuvD}), (\ref{eq:CorrelationDensity}) and (\ref{eq:Cuv}) are 2V densities, one could be upset with the superficially possible singular behavior of Eq. (\ref{eq:Gspec}) when simply taking the continuum limit without respecting the subtleties as shown in Appendix \ref{APP:2to1}.
To take the continuum limit, there may be many other ways which do not lead to the convergence to our results with continuum degrees of freedom or do not have any relevance to the discretization scheme. Some particularly strange limit was indicated in Appendix D of I; see Sec. \ref{conclusion} for further remarks. We will not discuss any of them in detail - the purpose of this paper is to present the scheme for bridging the continuum-degree density and the discretized one in a consistent way for both 2D and 3D (see below) gyrokinetics. The temperatures $\alpha(v)$ and $\alpha_0$ are determined by $G(v)$ and $E$ and should not change too much with the discretization scheme and resolution, and especially $\tilde{\alpha}(v_i)$ and $\tilde{\alpha}_0$ should converge to $\alpha(v_i)$ and $\alpha_0$.

\subsection{Partition function}
The gyrokinetic system was discretized from the beginning in I, however here we will start from the discretization of Eqs. (\ref{eq:H2Dc}, \ref{eq:CIuv}), as often done for the Feynman path integral. We will show how the same results can be obtained and how the continuum limit is recovered. We now denote $\tilde{\hat{g}}(\textbf{k},v_i)$ by $\hat{g}_{\textbf{k},i}$ for simplicity. Using Eq. (\ref{eq:delta}) to discretize Eqs. (\ref{eq:H2Dc}, \ref{eq:CIuv}), we get the Gibbs distribution, the same as in I,
\begin{eqnarray*}%\label{eq:disP}
\Pi_{\textbf{k}} \tilde{z}^{-1}_{\textbf{k}} \exp\Big{\{} \frac{1}{2} \sum_{i,j}[\delta_{ij} \alpha_i + \tilde{\alpha}_0 2 \pi \beta(\textbf{k})w_i(k) w_j(k)]\hat{g}_{\textbf{k},i}\hat{g}^{\ast}_{\textbf{k},j} \Big{\}},
\end{eqnarray*}
%\footnote{See Footnote \ref{fn:gki} for the notation of $\hat{g}_{\textbf{k},i}$, and Footnote \ref{realizability} for a factor of ``$2^N$''.}
%\cite{FT2n}
with the partition function for each $\textbf{k}$
\begin{eqnarray}\label{eq:PartitionFunction}
\tilde{z}_{\textbf{k}}=\pi^N \det\{ \delta_{ij} \alpha_i + \tilde{\alpha}_0 2 \pi \beta(\textbf{k})w_i(k) w_j(k) \}^{-1}.
\end{eqnarray}
[A factor of $2^N$ is eliminated by the realizability condition as in the previous subsection.]
The term $\delta_{ij} \alpha_i\hat{g}_{\textbf{k},i}\hat{g}^{\ast}_{\textbf{k},j}$ comes from $$\frac{\delta_{i,j}}{m_j}\tilde{\alpha}(v_i)\hat{g}_{\textbf{k},i}\hat{g}^{\ast}_{\textbf{k},j}m_i m_j \Leftarrow \delta(u-v)\alpha(v)\hat{g}(\textbf{k},u)\hat{g}^{\ast}(\textbf{k},v)dudv,$$ which exactly shows how the discretization scheme works.
Invoking the matrix determinant lemma,\cite{MDL} we have
\begin{eqnarray*}\label{eq:MDL}
\det\{ \delta_{ij} \alpha_i + \tilde{\alpha}_0 2 \pi \beta(\textbf{k})w_i(k) w_j(k) \}%\\
=[1+\sum_{i,j} \delta_{ij} \alpha^{-1}_i  \tilde{\alpha}_0 2 \pi \beta(\textbf{k})w_i(k) w_j(k)]\det\{\delta_{i,j}\alpha_i\}\\
=[1+\sum_{i,j} \delta_{ij} \alpha^{-1}_i  \tilde{\alpha}_0 2 \pi \beta(\textbf{k})w_i(k) w_j(k)]\Pi_i \alpha_i.
\end{eqnarray*}
Substituting this into Eq. (\ref{eq:PartitionFunction}), we get the partition function.
%\cite{FTnormalization}
%\footnote{
(It is possible to proceed from here to obtain the functional determinant and the functional partition function: Formal divergence may be met but shall not cause essential difficulty, as what we really need are the derivatives.)
%}
Then, we can derive the observables (by introducing the auxiliary function, if necessary). In particular, \begin{eqnarray}
\!\!\!\!\!\!\!\!\!  \langle \hat{g}_{\textbf{k},i}^{*} \hat{g}_{\textbf{k},i} \rangle \!\!=\!\! -\frac{\partial \ln \tilde{z}_{\textbf{k}}}{\partial \alpha_i} \!\! = \!\! \frac{1}{ \alpha_i} \left[ 1 \! - \! \frac{\tilde{\alpha}_0 2\pi
  \beta(k) w_i^2 \alpha_i^{-1}}{1 \! + \! \tilde{\alpha}_0 2\pi \beta(k)
  \sum_l w_l^2 \alpha_l^{-1}} \right]
\end{eqnarray}
which converges to the continuum result through the scheme for discretization/quantization given in the previous subsection. This equation is the same as in I. The nice thing is that $\langle \hat{g}_{\textbf{k},i}^{*} \hat{g}_{\textbf{k},i} \rangle$ is exactly conserved by the discretized dynamics as the starting point of I.

The above brief supplementary material thus consistently shows that the continuum calculation, if discretization is postpone to the intermediate stage, yields the same multivariable distribution as that from the discretized dynamics. Then, by an alternative method, the partition function, we obtain the same final results as I. These two approaches are conceptually different, but consistent. Especially, the way we obtain the discretized Gibbs distribution highlights the discretization scheme through which the results converge to the continuum results.

\section{Discussions}
\label{discussion}
For the physical discussions here, generally either 1V or 2V distribution will work: For the case with continuous velocity, obviously the 1V density is more convenient; for the quantized case with $m_i=\Delta v$, the corresponding 1V distribution is just Eq. (\ref{eq:Gspec}) multiplied by $\Delta v$ following Appendix \ref{APP:2to1}
%\cite{FTnoGi}
%\footnote{Earlier in I, $\mathrm{G}_i(\textbf{k})$ should have been used to denote the 1V discrete density instead of the 2V density $c_{i,i}(\textbf{k})$ in Eq. (11) there. For this historical reason, we resist using $G_i$ in this paper.}
so that we will just use the 2V density, Eq. (\ref{eq:Gspec}), without the necessity of extra work.

Eqs. (\ref{eqQNK}), (\ref{eq:Gvspectrum}) and (\ref{eq:elecspecC}) with negative $\alpha_0$ seem to argue for the dual-cascade scenario as suggested by Kraichnan in 2D Navier-Stokes turbulence.\cite{Kraichnan2D1967} The idea is simply that the system persists in relaxing to the absolute equilibrium state, with $E$ and $G$ concentrating more at, respectively, low- and high-$k$ modes (some minor oscillations may be introduced by $J_0$, but they should not change the picture essentially). However, the equilibrium state is continuously destroyed by the dissipation (, or other sink, and pumping).
When there is a minimal cutoff scale in velocity space as introduced by the discretization of velocity shown in Eq. (\ref{eq:Gspec}), an extra term emerges. As in the discussion of Appendix \ref{APP:2to1}, but with finite, instead of infinitesimal, $\Delta v$, the second term in Eq. (\ref{eq:Gspec}), $\frac{\tilde{\alpha}_0 2\pi \beta(k) w_i^2 \alpha_i^{-1}}{1+\tilde{\alpha}_0 2\pi \beta(k)\sum_l w_l^2 \alpha_l^{-1}}$, gives vanishing contribution to the distribution over $v$ of $G$ when $\Delta v\to 0$. New physics enter when the contribution from this second term is not negligible, which happens especially clearly in either of the following two cases where $\Delta v$ is not small enough and that the second term can even dominate: The denominator approaches zero from above with $\tilde{\alpha}_0<0$ and $\alpha_j>0$ with $j>0$, when $k\to k_{min}$; or, the denominator approaches zero from below with $\tilde{\alpha}_0>0$ and $\alpha_j<0$ with $j>0$, when $k\to k_{max}$. In general, we have $N+1$ conserved quantities (totally $N$ lattice points are introduced for the discretization of $g$ on $v$) defined by the summation over $\textbf{k}$ of Eqs. (\ref{eq:Gspec}) and the discretized version of (\ref{eq:elecspecC}), with their relations constrained by the discretized version of Eq. (\ref{eqQNK}). The spectral relaxation of all these $N+1$ conserved quantities with the nested constraints would be complicated and it is challenging to specify which case(s) is (are) physically relevant to turbulence (\textit{note that negative $\tilde{\alpha}(v_i)=\alpha_i/m_i$ is realizable}.) We can see that the discretized spectral relation between $D(\textbf{k})$ and $c_{i,i}(\textbf{k})$ could drastically differ from that of the continuous $v$ case when $\alpha_0$ is negative. Especially, when $k$ is small, say, well below some injection wavenumber $k_{in}$, the discretized version of Eq. (\ref{eq:elecspecC}), dominated by the second term, may have the same asymptotic spectral behavior as Eq. (\ref{eq:Gspec}). In such case the negative-$\alpha_0$ state with most of $E$ and $G$ located at the lowest modes indicates a possibility that all of them are transferred to lowest modes simultaneously as one whole conserved quantity.
%\cite{FTwz}
%\footnote{Such an absolute equilibrium state is realizable and has been verified in numerical simulations \cite{WatanabeZhu2011}.}
It is also possible that finite Larmor effect at intermediate scales may hinder the forward transfer of $G$ so that a large amount of $G$ may reside in the low mode regime. Suppose the wavenumber $k_{in}$ where both $E$ and $G$ are injected is some intermediate wavenumber, a possible picture would be that both $E$ and $G$ are transferred downward to ever larger scales but a forward transfer of $G$ dominates above $k_{in}$ to ever smaller scales where dissipation happens.
The finite $\Delta v$ introduces a particular truncation of velocity scale and can be related to collision/dissipation scale. The results also highlight the importance of sufficient numerical resolution in $v$ and the use of appropriate dissipation model or collision operator. The former is well known, but to our knowledge, this is the first time the resolution effects are calculated in the case of absolute equilibrium. For the latter, a particularly relevant and illuminating case from fluid turbulence was shown in Ref. \onlinecite{FrischPRL08}. The difference of a factor of $\Delta v$ between the 1V and 2V density spectra should also be taken into account in the evaluation of the pseudo-dissipation/collision from the thermalization of increasing modes during the transit regime of the simulations as done in fluids by Cichowlas et al.\cite{CBDBprl05}

With exactly the same ideas and techniques as in the calculations of the 2D case, but extending $v$ to $\vec{v}=(v_{\perp}, v_{\parallel})$ (c.f., the end of Appendix \ref{APP:FunctionalInverseII},) one can obtain the 3D results which are in consistent with Zhu and Hammett,\cite{ZhuHammettPoP2010} where there was no disagreement between the coauthors in taking the continuum limit and discussing the relevant physics.

\section{Concluding remarks}
\label{conclusion}

In summary, the exact gyrokinetic statistical absolute equilibria can be obtained analytically in two ways: either starting starting from the Galerkin truncated gyrokinetic equations or from their discretized versions (with respect to velocity field). That is, the results derived from the original integro-differential equations and those obtained by taking the continuum limit of the absolute equilibria of the discretized system are shown to be consistent. The comparison of the results brings physical insights such as possible variations of the generic feature of dual cascade in magnetized plasmas as a kind of kinetic effect. Most of the physical pictures discussed earlier in I with respect to the discretized results, persist in the continuous limit. Therefore, we have not repeated them here, but we have only isolated some specific issues.
%It thus provides the possibility of more systematic theories of kinetic plasma turbulence of magnetized plasmas.

The prescription of bridging the absolute equilibria between continuous velocity and quantized/discretized velocity,
$$\delta_{i,j}/m_i \! \leftrightarrows \! \delta(\!u-v\!); \alpha_i \!=\! \tilde{\alpha}(v_i)m_i; \tilde{\alpha}(v_i) \! \to \! \alpha(v_i); \tilde{\alpha}_0 \! \to \! \alpha_0,$$
should also be used for the evaluation of final equilibrium spectra with given initial $G(v)$ and $E$. Specific attention should be paid to the subtle differences between the 1V or 2V densities: For example, in Appendix D of I, $G_i$ was used for 1V densities and subsequently 2V ones in the right column of page 10, which may be the reason leading to the final unphysical conclusion, namely, in the continuum limit all the energy will be frozen in the lowest mode.
%\cite{FTapD}
%\footnote{
[Misconceptions there are evidenced by inconsistent conclusions which can be summarized as follows: For small $E$ and $\Delta v$, both $E \propto k$ and $E(k)=0 \ \text{for all} \ k>k_{min}$ were derived.]
%}
The prescription given here for regularization also means that more lattice sites (denoted by ``i'' in $G_i$ there) should appear with decreasing lattice spacing, which should have been taken into account in the end of Appendix D in I. $G_i$, evaluated at given velocities $(v_i,v_i)$, in the main text of I is conserved by the dynamics, which however does not mean that it should be fixed when taking the continuum limit: $G_i$s in Appendix D of I were fixed as $\Delta v \to 0$, then either all lattice points were squeezed together (in which case $i$ looses its meaning of velocity index) with the total grid number $N$ fixed or an infinite number of additional lattice points are inserted, in which case the information of the inserted $G$ is undefined. Neither case is physical. The unchanged quantity is the 1V density approximated by $G_i \Delta v$ for $\Delta v \to 0$ when $G$ is interpolated with given 1V density and with given (discretized) integral [summation of $G_i (\Delta v)^2$] of this 1V density over any velocity interval(s). This is similar to many other thermodynamic limit issues where the way of taking the limits is crucial to obtain correct or meaningful results.

This work also serves to outline a procedure for quantifying the errors due to finite truncation of velocity scale, although most physically realistic simulations
contain collision physics or other sources and/or sinks. The methodology may be used to explore other systems with similar features, such as other kinetic models of plasma dynamics, and it can be of interest to researchers concerned with developing turbulence model (numerical and analytical) for plasma descriptions based on the gyrokinetic ordering.

\begin{acknowledgements}
Part of the work was performed while the author was in the Theoretical Department of the Princeton Plasma Physics Laboratory through the University of Maryland with the supports of the U.S. Department of Energy Contract Nos. DE-FC02-04ER5478 and DE-AC02-09CH11466. The work in USTC was partially supported by the Fundamental Research Funds for the Central Universities. The final polishing of the manuscript also benefited from the support of Kavli Institute for Theoretical Physics China (KITPC) for the author's participation of the program ``New Directions in Turbulence,'' and from the support of the World Class Institute (WCI) Program of the National Research Foundation of Korea (NRF) funded by the Ministry of Education, Science and Technology of Korea (MEST) [WCI 2009-001]. Maxime Lesure and Dario Vincenzi kindly helped improve the English grammar. The author thanks Y. Camenen, D. Escande, G. W. Hammett, J. A. Krommes, G. Plunk, E. Startsev, R. Waltz, T.-H. Watanabe and T. Tatsuno for their interests and communications.
\end{acknowledgements}

\appendix
\section{Functional inversion}
Here we sketch the functional inversion problem and related calculations.

\subsection{\label{APP:FunctionalInverseI}A direct approach}

As $\int \delta(u-x)\delta(x-v)dx=\delta(u-v)$, we can write $C(\textbf{k},u,v) = \frac{ \delta(u-v)}{\alpha(v)}+\bar{C}(\textbf{k},u,v)$. Balancing the rest terms from Eq. (\ref{eq:CIC}), we have
\begin{eqnarray}\label{eq:Cbar}
&& \bar{C}(k,u,v) =\frac{\alpha_02\pi\beta(k)uvJ_0(ku)J_0(kv)}{\alpha(u)\alpha(v)} %\nonumber\\
 +\frac{1}{\alpha(u)}\int \alpha_02\pi \beta(k)uJ_0(ku)xJ_0(kx)\bar{C}(k,x,v)dx \nonumber\\
&& =\frac{\alpha_02\pi\beta(k)uvJ_0(ku)J_0(kv)}{\alpha(u)\alpha(v)} %\nonumber\\ \!\!\!\!
 +\frac{1}{\alpha(v)}\int \alpha_02\pi \beta(k)vJ_0(kv)xJ_0(kx)\bar{C}(k,u,x)dx.
\end{eqnarray}
The above formula demonstrates the symmetry of $\bar{C}(k,u,v)$ and that we can write $\bar{C}(k,u,v)=c(k,u)c(k,v)\bar{c}(k)$, which, combined with Eqs. (\ref{eq:CIC}, \ref{eq:Cbar}), gives the equation
\begin{eqnarray}\label{eq:factorizationCbarII}
\bar{c}(k)c(k,u)\Big{[} c(k,v) \!-\! \frac{vJ_0(kv)\alpha_0 2\pi \beta(k)}{\alpha(v)} \! \int \! xJ_0(kx)c(k,x)dx \Big{]} %\nonumber\\
=\frac{\alpha_0 2\pi \beta(k)uJ_0(ku)vJ_0(kv)}{\alpha(u)\alpha(v)} \nonumber
\end{eqnarray}
and a similar equation obtained by permuting $u$ and $v$. We determine (up to a constant which can be absorbed by $\bar{c}$) the form of $c$ as $c(k,u)=\frac{uJ_0(ku)}{\alpha(u)}$
and then determine $\bar{c}$, and finally $C$ as given in Eq. (\ref{eq:Cuv}).

\subsection{\label{APP:FunctionalInverseII}A general functional inversion formula}
Eq. (\ref{eq:Cuv}) can be straight forwardly obtained from the functional inversion formula,
\begin{eqnarray}\label{eq:FSM}
[A(u,v)+f(u)g(v)]^{\mathcal{I}}
= \int A^{\mathcal{I}}(u,x)[\delta(x-v)%\nonumber\\
+\int f(x)g(y)A^{\mathcal{I}}(y,v)dy]^{\mathcal{I}}dx \nonumber \\
= \int A^{\mathcal{I}}(u,x)\Big{\{}\delta(x-v)-\int f(x)g(y)A^{\mathcal{I}}(y,v)dy+\nonumber \\
\int \!\!\!\big{[} \!\!\!\int f(x)g(y)A(y,z)dy \int f(z)g(y)A^{\mathcal{I}}(y,v)dy \big{]} dz- ...\Big{\}} dx \nonumber \\
= A^{\mathcal{I}}(u,v)- \int A^{\mathcal{I}}(u,x)f(x)dx \int A^{\mathcal{I}}(x,v)g(x)dx %\nonumber\\
\big{\{} 1- \int \!\!\!\!\! \int f(y)A(x,y)g(x)dxdy+\nonumber \\
 + \int \!\!\!\!\! \int f(y)A(x,y)g(x)dxdy \int \!\!\!\!\! \int f(y)A(x,y)g(x)dxdy-...\big{\}} \nonumber \\
 =A^{\mathcal{I}}(u,v)- \frac{\int A^{\mathcal{I}}(u,x)f(x)dx\int A^{\mathcal{I}}(x,v)g(x)dx}{1+\int \int f(y)A^{\mathcal{I}}(x,y)g(x)dxdy}
\end{eqnarray}
the discretization of which is the Sherman-Morrison formula for matrix inversion.
In the first equality of the above formula, we have used $[\int A(u,w)B(w,v)dw]^{\mathcal{I}}= \int B^{\mathcal{I}}(u,w)A^{\mathcal{I}}(w,v)dw$ which can be verified directly.
The simple formal calculation to obtain Eq. (\ref{eq:FSM}) corresponds also to the direct extension of Ref. \onlinecite{BartlettAMS51}. To our knowledge, this functional inversion formula (where the denominator is of course assumed to be nonzero) in explicit form is new, though it can be considered as an extension of the already well known Sherman-Morrion formula for matrix inversion.

$u$ and $v$ in Formula (\ref{eq:FSM}) can be vectors in the $(3+2)$-D case.

\section{\label{APP:2to1}Reverting to 1V description}
In the main text, we have raised the V-dimension of $G$ in Eqs. (\ref{eq:H2Dc}, \ref{eq:CIuv}, \ref{eq:Cuv}) to 2V. We need to revert it to 1V.

No confusion should be caused as Eq. (\ref{eq:Cuv}) involves the distribution on $u$-$v$ plane while Eq. (\ref{eq:Gvspectrum}) is the distribution over the coordinate $v$.
Although such reverting to one dimensional distribution is intuitively direct, let us consider the following subtleties.
The results from Eqs. (\ref{eq:Cuv}) and (\ref{eq:CorrelationDensity}) can be understood by invoking the notion of infinitesimals.
What we should have is the covariance $\langle \hat{g}(\textbf{k},u)du\hat{g}^{\ast}(\textbf{k},v)dv \rangle$ and that we should call $\langle \hat{g}(\textbf{k},v)\hat{g}^{\ast}(\textbf{k},v) \rangle$ the self-correlation density. We have
\begin{equation}\label{eq:infinitesimal1D2D}
   \lim_{u \to v}\langle \hat{g}(\textbf{k},u)du\hat{g}^{\ast}(\textbf{k},v)dv \rangle={dv\over\alpha(v)}- \frac{2\pi\alpha_0 \beta(k) v^2 J_0^2( vk)} {\alpha^2(v)\!\! \left[ \! 1\!\!+\!\!2\pi\alpha_0 \beta(k)\!\! \int \!\! {x^2 \! J_0^2( kx)\over \alpha(x)}dx \! \right]}(dv)^2,
\end{equation}
where we have applied $\lim_{u \to v}\delta(v-u)du=1$, which can be easily understood with a particular nascent delta function $\eta_\epsilon(u-v) = \epsilon^{-1} [\text{sgn}(\epsilon/2-|u-v|)+1]/2\rightleftarrows \delta(u-v)$ to regularize the Dirac delta and $du \rightleftarrows \epsilon$, in the way consistent with the prescription for numerical discretization or for taking the continuum limit as shown in the main text. Eq. (\ref{eq:infinitesimal1D2D}) clearly shows that the 1V density is $1/\alpha(v)$ which, in nonstandard analysis, should be the standard part of the 1V density. [The corresponding 1V discrete distribution of the 2V discrete distribution $c_{i,j}$ can be easily obtained in the same way with simply the infinitesimals $du(v)$ changed to finite $\Delta u(v)$ and that an ``additional'' term survives from the second term of Eq. (\ref{eq:CuvD}) or (\ref{eq:Gspec}).]
The fact is that the second term has infinitesimal contribution to the infinitesimal (zigzagged) strip around the line $u=v$ in the $u$-$v$ plane, while the first term with Dirac delta function can be integrated throughout the strip to give the finite 1D distribution $1/\alpha(v)$: For any reasonably (well) behaved test function
$T(u,v)$,
\begin{eqnarray}\label{eq:int1D2D}
\!\!\!\!\!\!\!\!\!\!\!\int\int_{\lim u \to v} C(\textbf{k},u,v)T(u,v)dudv %\nonumber\\
 =\lim_{\Delta u \to 0}\int dv \int_{v-\Delta u/2}^{v+\Delta u/2}[\frac{ \delta(u-v)}{\alpha(v)}+\bar{C}(\textbf{k},u,v)]T(u,v)du \nonumber \\
\!\!\!\!\!\!\!\!\!\!\!=\int \frac{1}{\alpha(v)}T(v,v)dv %+ \nonumber\\
+\lim_{\Delta u \to 0}\int \{\bar{C}(\textbf{k},v,v)T(v,v)\Delta u +O[(\Delta u)^2]\}dv %\nonumber \\
=\int \frac{1}{\alpha(v)}T(v,v)dv.
\end{eqnarray}
(Interchangability of integration and taking limit is assumed in the above.)

\end{CJK*}
\end{CJK*}


\begin{thebibliography}{10}

\bibitem[Lee(1952)]{Lee1952}
T.-D. Lee, {\em Q. Appl. Math.} {\bf 10}
69 (1952)% On some statistical properties of hydrodynamic and hydromagnetic fields. .

\bibitem[Krstulovic \& Brachet (2011)]{KrstulovicBrachet2011}
G. Krstulovic and M. Brachet, {\em Phys. Rev. E} {\bf 83} 066311 (2011).% Energy cascade with small-scale thermalization, counter?ow metastability, and anomalous velocity of vortex rings in Fourier-truncated Gross-Pitaevskii equation. .

\bibitem[Zhu \& Hammett (2010)]{ZhuHammettPoP2010}
J.-Z. Zhu and G. W. Hammett, {\em Phys. Plasmas} {\bf 17}, 122307 (2010).%Gyrokinetic statistical absolute equilibrium and turbulence.

%\bibitem{FTref}
%A minimization of plasma physics references is thus managed to keep the general readers in an easy environment.

\bibitem[Krommes(2012)]{Krommes12}
J. A. Krommes, {\em Annual Review of Fluid Mechanics\/} {\bf 44}, to be published (2012).% The Gyrokinetic Description of Microturbulence in Magnetized Plasmas. .

\bibitem[L'vov et al.(2002)]{LPPprl02}
V. S. L'vov, A. Pomyalov and I. Procaccia, {\em Phys. Rev. Lett.} {\bf 89} 064501 (2002). % Quasi-Gaussian Statisitcs of Hydrodynamic Turbulence in $4/3+\epsilon$ Dimensions. .

\bibitem[Frisch et al.(2008)]{FrischPRL08}
U. Frisch et al., {\em Phys. Rev. Lett.} {\bf 101} 144501 (2008).% Gakerkin Truncation, Hyperviscoisty and Bottleneck in Turbulence .

%\bibitem[Brizard \& Hahm (2007)]{BrizardHahmRMP07}
%A. Brizard and T.-S. Hahm, {\em Rev. Mod. Phys.} {\bf 79} 421 (2007).% Foundations of Modern Gyrokinetics. .

\bibitem[Smagorinsky (1963)]{Smagorinsky63}
The explicit spacial integration form however reminds us the large eddy simulation proposed by J. Smagorinsky, {\em Monthly Weather Review} {\bf 91} 99 (1963 ). %General Circulation Experiments with the Primitive Equations. .

%\bibitem{FTre}
%Writing $f=f-g+g$ and assuming $f-g$ is also a probability distribution function (which is true as long as the average of $g$ is zero) we can also directly see that the (information-theoretic) \textit{relative} entropy $-\int\!\!\!\int f\ln[f/(f-g)] d^2\textbf{r}dv = -\frac{1}{2}\int\!\!\!\int g^2/(f-g)+O(g^3)d^2\textbf{r}dv$.

\bibitem[Escande (1994)]{EscandeJSP94}
There does not seem to be a complete theory for applying such a distribution [but see, for example, D. Escande et al., {\em J. Stat. Phys.} {\bf 76} 605 (1994), for a validity check of the Gibbs technique; and, see A. I. Khinchin, Mathematical Foundations of Statistical Mechanics, Dover Publications, (1949) for some fundamental supports and cautions.] % Self-consistent check of the validity of Gibbs calcus using dynamical variables.

%\bibitem{FieldIntegral}
%In quantum and statistical field theory [See, e.g., A. Altland and B. Simons, Condensed Matter Field Theory, second edition, Cambridge University Press (2010); and, A. Das, Field Theory: A Path Integral Approach, World Scinetific (1993),] $C^{\mathcal{I}}$ here is called the {\it operator} kernel or {\it propagator} and $C$ below can be interpreted as the {\it Green function} of $C^{\mathcal{I}}$. [The measure $D\sigma$ is usually written as $D(\hat{g}^{\ast},\hat{g})$ with $\hat{g}$ being the function of $\textbf{k}$ and $x$. But, note that our stationary integration measure does not depend on time $t$.] Our case is just the functional Gaussian (but not Grassmann) integral, but the propagator presents new features which require novel techniques.

%\bibitem{FTdv}
%``D'' is already used to denote the dimension in configuration space, so we adopt ``V'' here for the dimension in velocity space (V-dimension.)

%\bibitem{FTrealizability}
%A factor of ``2'' has been eliminated to respect the realizability condition $g^*(\textbf{k},u)=g(-\textbf{k},u)$.

%\bibitem{FTdis}
%In I, the same covariance density matrix as in Eq. (\ref{eq:CuvD}) was obtained by starting with the discretization of Eqs. (\ref{eq:2DgkK}) and (\ref{eqQNK}). Especially, to facilitate the calculation, I multiplied $\tilde{G}_i$ by $\delta_{i,j}$, which, according to Eq. (\ref{eq:delta}), transformed the 1V density to 2V density: $\tilde{G}_i/m_i \Rightarrow G_i$, with the notations used there.

\bibitem[Mandelbrot(1974)]{MandelbrotJFM74} B. B. Mandelbrot, {\em J. Fluid Mech.} {\bf 62} 331 (1974).% Intermittent turbulence in self-similar cascades: divergence of high moments and dimension of carrier. .

%\bibitem{FTfractal}
%Note however that we have not proposed here, in velocity space, the fractals which may be appropriate for some physical properties though. With the fractal as an example, it is indicated that the technique may be extended to cases with fractional dimensions.

%\bibitem{FTgi}
%$\mathrm{G}_i$ in Eq. (11) of I is a 2V density so that, even though the factor of $1/2$ there is neglected, the summation over $\textbf{k}$ of it does not equal to $\tilde{G}_i$, which is 1V, in the left column on that same page.

\bibitem[Watanabe \& Zhu(2011)]{WatanabeZhu2011}
T.-H. Watanabe, private communication (2011).

%\bibitem{FTgki}
%So, to be accurate, for the discretized case earlier in I we should have replaced $\hat{g}(\textbf{k},v_i)$ with $\tilde{\hat{g}}(\textbf{k},v_i)$, denoted by $\hat{g}_{\textbf{k},i}$ for clarity, as is the case of distinguishing the solution of the discretized equation (say, the difference equation) of a differential equation and the solution of the latter.

%\bibitem{FT2n}
%See Footnote \onlinecite{FTgki} for the notation of $\hat{g}_{\textbf{k},i}$, and Footnote \onlinecite{FTrealizability} for a factor of ``$2^N$''.

\bibitem[Harville(1997)]{MDL}
D. A. Harville, Matrix Algebra From a Statistician's Perspective. Springer-Verlag (1997).

%\bibitem{FTnormalization}
%It is possible to proceed from here to obtain the functional determinant and the functional partition function (formal divergence may be met but shall not cause essential difficulty, as what we really need is the derivatives,) but we don't bother to do it now.

%\bibitem{FTnoGi}
%Earlier in I, $\mathrm{G}_i(\textbf{k})$ should have been used to denote the 1V discrete density instead of the 2V density $c_{i,i}(\textbf{k})$ in Eq. (11) there. For this historical reason, we resist using $G_i$ in this paper.

\bibitem[Kraichnan(1967)]{Kraichnan2D1967}
R.~H. Kraichnan, {\em
  Phys. Fluids\/} {\bf 102}, 1417 (1967).% Inertial ranges in two-dimensional turbulence. .

%\bibitem{FTwz}
%Earlier in I, $\mathrm{G}_i(\textbf{k})$ should have been used to denote the 1V discrete density instead of the 2V density $c_{i,i}(\textbf{k})$ in Eq. (11) there. For this historical reason, we resist using $G_i$ in this paper.

\bibitem{CBDBprl05}
C. Cichowlas et al., {\em Phys. Rev. Lett.\/} {\bf 95}, 264502 (2005).

%\bibitem{FT3d}
%This paragraph is only for readers interested in 3D continuum-limit gyrokinetic absolute equilibrium on which, we, the authors of I have no disagreement and which is not the focus of this work.

%\bibitem{FTapD}
%Misconceptions there are evidenced by those inconsistent conclusions: For small $E$ and $\Delta v$, both $E \propto k$ and $E(k)=0 \ \text{for all} \ k>k_{min}$ were derived.

\bibitem[Bartlett(1951)]{BartlettAMS51}
M. S. Bartlett, {\em Annals of Mathematical Statistics}
{\bf 22} 107 (1951).% An Inverse Matrix Adjustment Arising in Discriminant Analysis. ).





\end{thebibliography}
\end{document}